\documentclass[fleqn,twoside]{article}
\usepackage{espcrc2}
\usepackage{epsf}

\def\prd#1{Phys.\ Rev.\ {\bf D#1}}

\title{Heavy-light decay constants using clover valence quarks and three
flavors of dynamical improved staggered quarks}
                                                
\author{The MILC Collaboration: C.~Bernard
\address{Department of Physics, Washington University, St.~Louis, MO 63130, USA},
S. Datta
\address{Fakult\"at f\"ur Physik, Universit\"at Bielefeld, D-33615 Bielefeld, Germany},
C.~DeTar
\address{Physics Department, University of Utah, Salt Lake City, UT 84112, USA},
Steven Gottlieb
\address{Department of Physics, Indiana University, Bloomington, IN 47405, USA}\thanks{Presented by S.~Gottlieb},
E.B.~Gregory
\address{Department of Physics, University of Arizona, Tucson, AZ 85721, USA},
U.M.~Heller
\address{American Physical Society, One Research Road, Ridge, NY 11961--9000, USA},
J.~Osborn$\,\null^{\rm c}$,
R.L.~Sugar
\address{Department of Physics, University of California, Santa Barbara, CA 93106, USA} and
D.~Toussaint$\,\null^{\rm e}$
}

\begin{document}

\begin{abstract}
Starting in 2001, the MILC Collaboration began a large scale
calculation of heavy-light meson decay constants using clover valence
quarks on ensembles of three flavor configurations.  For the coarse
configurations, with $a=0.12$ fm, eight combinations of dynamical light and
strange quarks have been analyzed.  
For the fine configurations, with $a=0.09$ fm,
three combinations of quark masses are studied.  Since we last reported
on this calculation, statistics have been increased on the fine ensembles,
and, more importantly, a preliminary value for the perturbative 
renormalization of the axial-vector current has become available.
Thus, results for $f_B$, $f_{B_s}$, $f_D$ and $f_{D_s}$ can, in principle, 
be calculated in MeV, in addition to decay-constant ratios that were 
calculated previously.

\end{abstract}

\maketitle

\section{INTRODUCTION}

Calculation of heavy-light meson decay constants is important for the 
extraction of a number of CKM matrix elements.  For example, 
 $\Gamma(D^+_s\rightarrow l^+\nu_l) \propto f^2_{D_s} |V_{cs}|^2$.  
As decay constants are fairly easy to determine in lattice calculations
and they will be measured at BaBar, KEK and CLEO-c,
they provide an excellent opportunity to verify the accuracy of our methods.

We are extending a calculation of heavy-light meson decay constants
with three flavors of dynamical quarks that was begun in 2001 \cite{LAT01}.
In this calculation, clover quarks are used for both the light and 
heavy valence quarks, the latter with the Fermilab interpretation \cite{EKM}.
A collaboration of Fermilab
and MILC is now using light Asqtad quarks with the same dynamical gauge
configurations to calculate decay constants.  The newer calculation
\cite{SIMONE} will
allow better control of chiral extrapolations, as it is practical to reduce the
light valence quark mass.  However, for $D_s$ and $B_s$ we expect
the two approaches to be complimentary.

Since the last time we reported on this calculation \cite{LAT02}, 
we have greatly increased the statistics on two gauge ensembles 
with $a=0.09$ fm.  (See Table 1.)  Furthermore,
preliminary results from a one-loop perturbative calculation of the
axial-vector renormalization constant $Z_A$ have been provided 
by El-Khadra, Nobes and Trottier \cite{NOBES}.

\begin{table}
\begin{tabular}{cccc}
\noalign{\hrule}
\small
  dynamical & $\beta$ & configs.    & configs.  \\
\noalign{\vskip -.07truein}
              $am_{u,d}/am_s$ &      & generated   & analyzed  \\
\noalign{\hrule}
\noalign{\hrule}
\noalign{\smallskip}
\noalign{\hrule}
\noalign{\hrule}
\noalign{\smallskip}
\multicolumn{4}{c}{$a=0.09$ fm; $28^3\times96$}\\
\noalign{\hrule}
\noalign{\smallskip}
    0.031/0.031   & 7.18 & 496 &163 (163) \\   
    0.0124/0.031   & 7.11 & 527 &{  242} (120) \\ 
    0.0062/0.031   & 7.09 & 592  &{ 293} (48)\\ 
\noalign{\hrule}

\end{tabular}
\caption{Details of fine lattice configurations.  In parentheses are 
the numbers of configurations analyzed in Ref.~\cite{LAT02}.
}
\end{table}

Dynamical gauge configurations are generated using the Asqtad action 
\cite{ASQTAD}.  Further details of confiuration generation may be found in
Ref.~\cite{MILCSPEC}.
For each ensemble of dynamical quark configurations,
we use five light and five heavy valence quark masses.  The masses and
decay constants are interpolated or extrapolated as explained below to 
get physically relevant values.  The relative scale is set through
the heavy quark potential \cite{MILCSPEC}
and the overall scale is set from bottomonium splittings.

\section{ANALYSIS OF RESULTS}


The analysis of the heavy-light decay constants involves a number of steps.
On each ensemble studied, we:
\begin{enumerate}
\item{fit light pseudoscalar (PS) hadron propagators to determine PS masses}
\item{perform a quadratic chiral fit of squared PS masses to determine $\kappa_c$ 
}


\item{determine $m_s$ from the mass of $\bar s s$ pseudoscalar state assuming
a linear chiral mass relation}
\item{fit heavy-light (HL) channels to determine their masses and decay amplitudes
}


\item{extrapolate or interpolate results in light quark mass to $m_{u,d}$
or $m_s$, respectively
(see Fig.~1).  
We see that the interpolation required for the strange quark mass is well
under control.  However, in extrapolating to $m_{u,d}$, we are expected to
find chiral logarithms \cite{CHIRALLOGS}.  With clover light quarks, it is a
long extrapolation to the physical value of $m_{u,d}$ and it is not 
possible to see the chiral logs.  This difficulty is much improved with Asqtad
light quarks \cite{SIMONE}.
We could also consider an extrapolation of decay constant ratios, such as
$f_B/f_\pi$ or $(f_{B_s}/f_B)/(f_K/f_\pi)$ \cite{RATIOEXTRAP}.
}

\begin{figure}[t]
\epsfxsize=0.99 \hsize
\epsffile{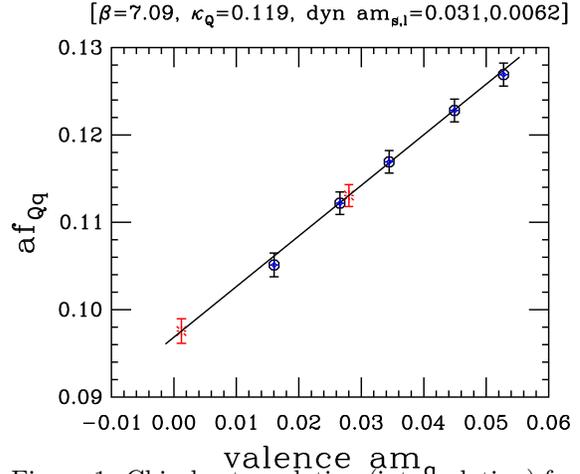}
\vspace{-38pt}
\caption{Chiral extrapolation (interpolation) for $f_B$ ($f_{B_s}$)
for $\beta=7.09$, $\kappa_Q=0.119$.}
\end{figure}

\item{after removal of perturbative logarithms, fit $f_{Qq}\sqrt{M_{Qq}}$ 
to a power series in $1/M_{Qq}$ and interpolate to $B$, $B_s$, $D$ and $D_s$ 
meson masses
}


\item{put the perturbative logarithm back and use 
the heavy-light axial-vector current
renormalization constant to get the renormalized decay-constant}
\end{enumerate}

Just before the conference, we obtained the 
preliminary results of a perturbative calculation
of the axial-vector renormalization constant $Z_A$ \cite{NOBES}.
As these results are preliminary, and our use of the results has not been
as thoroughly checked as we would like, the following results are to be
considered preliminary.  In particular, we have not yet tried tadpole 
improvement, which should help to determine the size of our systematic error.
We note that 
when we reported on this calculation at Lattice 2002 \cite{LAT02}, 
no perturbative or nonperturbative calculation of $Z_A$ was available.
We used an ad hoc procedure based on comparison of our improved action
quenched results with the continuum limit of earlier calculations using the
Wilson gauge action and Wilson or Clover quarks.
This was explained in more detail in Ref.~\cite{LAT01}.

After steps 1--7 are completed on each ensemble, we have a partially
quenched result at a particular value of dynamical $m_\pi/m_\rho$.
We then plot these results as a function of $(m_\pi/m_\rho)^2$ to perform
a chiral extrapolation for the sea quarks.  
Before looking at the decay constant of a specific meson, 
we note that if we plot the ratio of
decay constants for two different mesons a good deal of the
uncertainty from the renormalization constants and 
other systematic errors drops out.
Figure 2 shows the ratio $f_{B_s}/f_B$ of meson decay constants.
In Fig. 3, we show $f_{B_s}$.

\begin{figure}[t]
\epsfxsize=0.99 \hsize
\epsffile{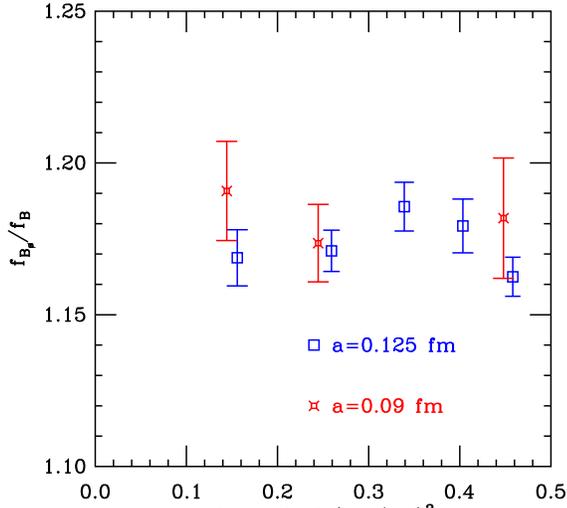}
\vspace{-38pt}
\caption{$f_{B_s}/f_B$ as a function
of $(m_\pi/m_\rho)^2$.}
\end{figure}

\begin{figure}[t]
\epsfxsize=0.99 \hsize
\epsffile{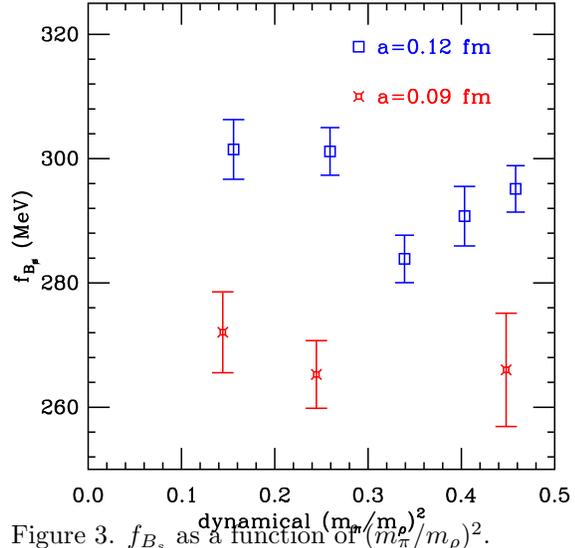}
\vspace{-38pt}
\caption{$f_{B_s}$ as a function of $(m_\pi/m_\rho)^2$.}
\end{figure}


\section{FUTURE WORK}

We need to complete this analysis by including alternative cuts on the
fits of meson propagators and alternative chiral extrapolations.  We also must
check the effect of tadpole improvement and see if we can employ a trick of
Ref.~\cite{TRICK}  in which $Z_A^{hl} = \rho_A^{hl} \sqrt{Z_V^{hh}Z_V^{ll}}$
where $\rho_A^{hl}$ is computed perturbatively and both $Z_V$'s 
nonperturbatively.
Although it would be possible to increase statistics on the fine configurations
or include newer coarse ensembles ($a=0.12$ fm), our more recent effort 
with Asqtad light quarks appears to be more promising.

\end{document}